%% file: prl_shortened.tex
\documentclass[prl,twocolumn,showpacs,superscriptaddress,floatfix]{revtex4}
\usepackage{epsfig}
\usepackage{graphicx}
\usepackage{dcolumn}
\usepackage{amsmath}
\usepackage{xspace}
\newcommand{\pbarp}{{\bar p}p}
\newcommand{\roots}{{\sqrt s}}
 
\newcommand{\Et}{E_T}
\newcommand{\Pt}{p_T}
\newcommand{\Ht}{H_T}

\newcommand{\gt}{>}
\newcommand{\lt}{<}

\newcommand{\mett}{{\not\!\!E}_{T}}

\newcommand{\coriso}{E^{iso}_{cone}}

\newcommand{\MeV}{\ensuremath{\mathrm{\ Me\kern -0.1em V}}\xspace}
\newcommand{\MeVc}{\ensuremath{\mathrm{\ Me\kern -0.1em V\kern -0.1em 
/\mathit{c}}}\xspace}
\newcommand{\MeVcsq}{\ensuremath{\mathrm{\ Me\kern -0.1em V\kern -0.1em 
/\mathit{c}^2}}\xspace}
\newcommand{\GeV}{\ensuremath{\mathrm{\ Ge\kern -0.1em V}}\xspace}
\newcommand{\GeVc}{\ensuremath{\mathrm{\ Ge\kern -0.1em V\kern -0.1em 
/\mathit{c}}}\xspace}
\newcommand{\GeVcsq}{\ensuremath{\mathrm{\ Ge\kern -0.1em V\kern -0.1em 
/\mathit{c}^2}}\xspace}
\newcommand{\TeV}{\ensuremath{\mathrm{\ Te\kern -0.1em V}}\xspace}
\newcommand{\nsGeV}{\ensuremath{\mathrm{Ge\kern -0.1em V}}\xspace}
\newcommand{\nsGeVc}{\ensuremath{\mathrm{Ge\kern -0.1em V\kern -0.1em 
/\mathit{c}}}\xspace}
\newcommand{\Etgamma}{\ensuremath{\mathrm{E_T^{\gamma}}}}
\newcommand{\Etlepton}{\ensuremath{\mathrm{E_T^{\ell}}}}
\begin{document}

\preprint{FERMILAB-PUB-02/031-E}
\preprint{EFI-02-66}

\title{Search for New Physics in Photon-Lepton Events in $p{\bar p}$
Collisions at $\roots= 1.8\TeV$}


\input{run1_revtex4_auth.tex}
\date{\today}

\begin{abstract}
We present the results of a search in $p{\bar p}$ collisions at
$\roots= 1.8\TeV$ for anomalous production of events containing a
photon and a lepton ($e$ or $\mu$), both with
large transverse energy, using 86 pb$^{-1}$ of data collected with the
Collider Detector at Fermilab during the 1994-95 collider run at the
Fermilab Tevatron.  The presence of large missing transverse energy
($\mett$), additional photons, or additional leptons in these events
is also analyzed.  The results are consistent with standard model
expectations, with the possible exception of photon-lepton events with
large $\mett$, for which the observed total is 16 events and the
expected mean total is $7.6\pm0.7$ events.
\end{abstract}

\pacs{13.85.Rm, 12.60.Jv, 13.85.Qk, 14.80.Ly}

\maketitle

An important test of the standard model (SM) of particle
physics~\cite{SM} is to measure and understand the properties of the
highest-energy particle collisions.  The observation of an anomalous
production rate of any combination of the fundamental particles 
of the SM would be a clear indication of a new physical
process.  This Letter summarizes an analysis of the inclusive
production of a photon and a lepton ($e$ or $\mu +\gamma + X$),
including searches for additional
photons, leptons, and large missing transverse energy, using
86~pb$^{-1}$ of data from proton-antiproton collisions collected with
the Collider Detector at Fermilab (CDF) during the 1994-95 run of the
Fermilab Tevatron~\cite{jeffPRD}.

Production of these particular combinations of particles is of
interest for several reasons.  Events with photons and leptons are
potentially related to the puzzling ``$ee\gamma\gamma\mett$'' event
recorded by CDF~\cite{eegg}.  A supersymmetric model~\cite{susy}
designed to explain the $ee\gamma\gamma\mett$ event predicts the
production of photons from the radiative decay of the
$\tilde{\chi}^{0}_{2}$ neutralino, and leptons through the decay of
charginos, indicating $\ell\gamma\mett$ events as a signal for the
production of a chargino-neutralino pair.  Other hypothetical, massive
particles could subsequently decay to SM electroweak gauge bosons, one
of which could be a photon and the other a $W$ or $Z^0$ boson that
decays leptonically.  In addition, photon-lepton studies complement
similarly-motivated inclusive searches for new physics in
diphoton~\cite{gg}, photon-jet~\cite{gj}, and photon-$b$-quark
events~\cite{gb}.

The CDF detector~\cite{cdfblurb} is a cylindrically symmetric
spectrometer designed to study $\pbarp$ collisions at the Fermilab
Tevatron.  A superconducting solenoid of length 4.8~m and radius 1.5~m
generates a magnetic field of 1.4~T and contains tracking chambers
used to measure the momenta of charged particles.  A set of vertex
time projection chambers is used to find the $z$
position~\cite{CDFcoo} of the $\pbarp$ interaction. The 3.5-m-long
central tracking chamber (CTC) is a wire drift chamber which provides
up to 84 measurements between the radii of 31.0~cm and 132.5~cm in the
region $|\eta| < 1.0$.  Sampling calorimeters, used to measure the
electromagnetic and hadronic energy deposited by electrons, photons,
and jets of hadrons, surround the solenoid. 
Each tower of
the central ($|\eta| < 1.1$) electromagnetic calorimeter (CEM) has
an embedded strip chamber for the measurement of
the 2-D transverse profile of electromagnetic showers.
Muons are detected with three systems of muon chambers, each
consisting of four layers of drift chambers.  The central muon
(CMU) system is located directly outside the central
hadronic calorimeter, and covers $|\eta| < 0.6$.  Outside of the CMU
is 0.6~m of steel shielding, followed by the central muon
upgrade system.  The central muon extension  system
provides muon detection for $0.6<|\eta|<1.0$.

Events with a high-transverse momentum ($\Pt$)~\cite{EtPt} photon or
lepton are selected by a three-level trigger~\cite{jeffPRD}, which
requires an event to have either a high-$\Et$ photon or a
high-$\Pt$ lepton ($e$ or $\mu$) within the central region, $|\eta| <
1.0$. Photon and electron candidates are chosen from clusters of
energy in adjacent CEM towers; electrons are then further separated
from photons by requiring the presence of a CTC track pointing at the
cluster. Muons are identified by requiring CTC tracks to extrapolate
to a reconstructed track segment in the muon drift chambers.

To reduce background from
the decays of hadrons produced in jets, both the photon and the lepton
in each event are required to be `isolated'. The $\Et$
deposited in the calorimeters in a cone in $\eta-\phi$ space of radius
$R=0.4$ around the photon or lepton position is summed, and the $\Et$
due to the photon or lepton is subtracted.  The remaining $\Et$ 
in the cone, $\coriso$, is required to be less than 2\GeV for a
photon, or less than 10\% of the lepton transverse momentum.  In
addition, for photons the sum of the $\Pt$ of all
tracks in the cone must be less than 5\GeV.

Inclusive photon-lepton events are selected by requiring
an isolated central photon with $\Etgamma>25$ GeV and an isolated
central lepton ($e$ or $\mu$) with $\Etlepton>25$ GeV. The
technical criteria used to identify leptons and photons are very
similar to those of Refs.~\cite{gg,gj,gb}, and are described in detail
in Ref.~\cite{jeffPRD}. 
A total of 77 events pass this selection: 29
photon-muon and 48 photon-electron candidates.

The production of pairs of new heavy states that decay via cascade decays
can lead to final states with multiple
photons or leptons; in contrast, the dominant SM background processes
lead to signatures with  only one photon and one lepton observed in
the detector, as discussed in detail below.
The inclusive sample is consequently analyzed as two subsamples: a
``two-body inclusive photon-lepton sample'' typical of a two-particle
final state, and a ``multi-body inclusive photon-lepton sample''
typical of three or more particles in the final state.  The two-body
sample selection requires exactly one photon and exactly one lepton,
with an azimuthal separation $\Delta
\varphi_{\ell\gamma}> 150^{\circ}$, but  excludes 
those events for which the invariant mass of the photon and electron,
$M_{e\gamma}$, is within 5\GeV of the mass of the Z boson, $M_Z$ (these events are used as a
control sample, as described below).  The multi-body sample is
composed of the remaining inclusive photon-lepton events.  The
multi-body sample is then further analyzed for the presence of large
$\mett$, and additional isolated leptons and photons.  The $\mett$ threshold of
25\GeV was determined \textit{a priori} in previous
analyses~\cite{r_papers} as a significant indicator of a neutrino
arising from leptonic decays of the $W$ boson. Figure~\ref{lgamma
path} shows the breakdown of the inclusive sample into the final
categories.

\begin{figure}[!ht]
\begin{center}
\includegraphics*[width=0.5\textwidth]{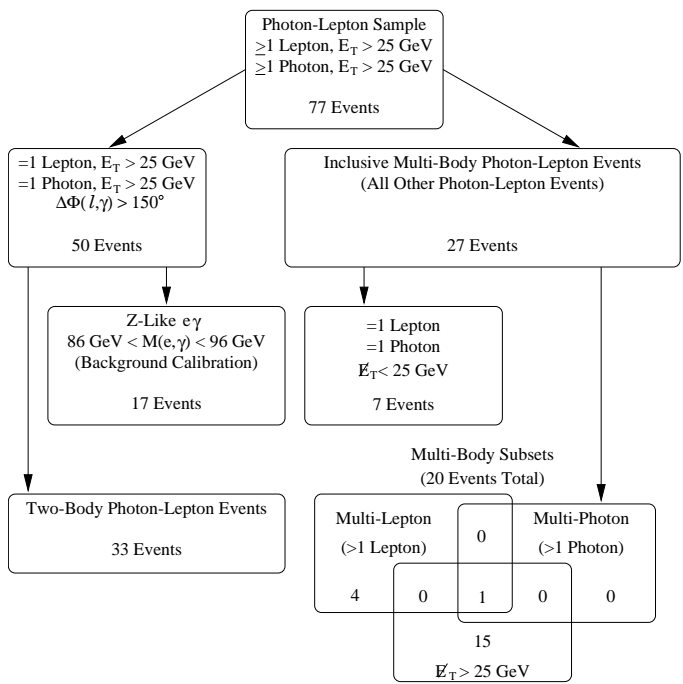}
\end{center}
\caption{The subsets of 
inclusive photon-lepton events.  The multi-body photon-lepton
subcategories of $\ell\gamma\mett$, multi-lepton, and multi-photon
events are not mutually exclusive.}
\label{lgamma path}
\end{figure}

The dominant source of photon-lepton events at the Tevatron is
electroweak diboson production, in which a $W$ or $Z^0$ boson decays
leptonically ($\ell \nu$ or $\ell\ell$) and a photon is radiated from
either an initial-state quark, a $W$, or a charged final-state lepton.
The number of such events is estimated using leading-order (LO) matrix
element calculations~\cite{diboson matrix} for which the computational
code~\cite{diboson mc} was then embedded into the general-purpose
event generator program \textsc{pythia}~\cite{pythia}, followed by a
full simulation of the detector.  The uncertainty in this number has
roughly equal contributions from higher-order processes, simulation
systematics, luminosity, proton structure, and generator statistics.

A jet can contain mesons such
as the $\pi^0$ or $\eta$ that decay to photons, which then
may satisfy the photon selection criteria.  The number of
lepton-plus-misidentified-jet events is determined by counting the
number of jets in a sample of events with a lepton 
and then multiplying by the probability
of a jet being misidentified as a photon, $P^{jet}_{\gamma}$.  The
factor $P^{jet}_{\gamma}$ is determined from samples of jets and
photons in events with a lepton trigger, using the distribution in
$\coriso$.  By fitting to a sum of the expected distribution for
prompt photons and that measured for jets,
the misidentification rate is found to be
$P^{jet}_{\gamma}=(3.8\pm0.7)\times10^{-4}$~\cite{jeffPRD}.

The dominant source of misidentified photon-electron events is
$Z^{0}\rightarrow e^+e^-$ production, where one of the electrons
undergoes hard photon bremsstrahlung or a track fails to be
reconstructed.  We assume that photon-electron events consistent with
$Z^{0}$ production are not a significant source of new physics, and
use them to estimate the probability $P^{e}_{\gamma}$ that an electron
is reconstructed as a photon. The number of misidentified
photon-electron events in the control sample
divided by the number of electron-electron events with the same
kinematics gives $P^{e}_{\gamma}= (1.28\pm0.35)$\%.
For any other subset of central electron pairs, the contribution to
the corresponding photon-electron sample is the product of
$P^{e}_{\gamma}$ and the number of central electron pairs.

Other, smaller, backgrounds are due to hadrons faking muons, and to
leptons from the decay of bottom and charm quarks.  Charged hadrons 
may penetrate the
calorimeters into the muon chambers, or may decay to a muon
before reaching the calorimeters.  These contributions are determined
by identifying isolated, high-momentum tracks in the inclusive photon
sample, applying the probability of each track of being misidentified
as a muon, and summing this probability over all tracks in the
sample~\cite{jeffPRD}.  The contribution to photon-lepton candidates
from heavy-flavor produced in association with a prompt photon is
estimated using Monte Carlo event generation~\cite{pythia} and
detector simulation, and found to be negligible.

New physics in small samples of events would most likely manifest
itself as an excess of observed events over expected events.  
The significance of an
observed excess is computed from the likelihood of obtaining at least the
observed number of events, $N_{0}$, assuming that the null hypothesis
(the SM) is correct.  The ``observation likelihood'', $P(N\geq
N_{0}|\mu_{SM})$, is defined as the fraction of the Poisson
distribution for the number of expected events from SM sources, with a
mean $\mu_{SM}$, that yields outcomes $N \ge N_{0}$~\cite{delta_mu}. 
The likelihood $P(N\geq N_{0}|\mu_{SM})$
is computed from a large ensemble of calculations in which each
quantity used to compute photon-lepton event sources varies randomly
as a Gaussian distribution, and the resulting mean event total is used
to randomly generate a Poisson-distributed outcome $N$.  The fraction
of calculations in the ensemble with outcomes $N \geq N_{0}$ gives
$P(N\geq N_{0}|\mu_{SM})$.

The predicted and observed totals for two-body photon-lepton events
are compared in Table~\ref{lg result}.  Half of the predicted total
originates from $Z^{0}\gamma$ production where one of the charged
leptons has escaped identification; the other half originates from
roughly equal contributions of $W\gamma$ production, misidentified
jets, misidentified electrons, and misidentified charged hadrons.  The
likelihood of the observed total is 9.3\%.
 
\begin{table}[!t]
\begin{center}
\begin{tabular}{lrrrr}\hline\hline
Process & 
\multicolumn{1}{c}{Two-Body} & 
\multicolumn{3}{c}{Multi-Body} \\
& 
\multicolumn{1}{c}{$\ell\gamma X$} & 
\multicolumn{1}{c}{$\ell\gamma X$} & 
\multicolumn{1}{c}{$\ell\gamma\mett X$} &
\multicolumn{1}{c}{$\ell\ell\gamma X$} 
\\ \hline
W$+\gamma$& 
$2.7\pm0.3$ & $5.0\pm0.6$ & $3.9\pm0.5$&\multicolumn{1}{c}{--} \\
Z$+\gamma$&
$12.5\pm1.2$ & $9.6\pm0.9$ &$1.3\pm0.2$& $5.5\pm0.6$\\ 
$\ell$+jet, jet $\rightarrow \gamma$ &
$3.3\pm0.7$ & $3.2\pm0.6$ &$2.1\pm0.4$& $0.3\pm0.1$\\ 
$Z \rightarrow ee, e \rightarrow \gamma$  &
$4.1\pm1.1$ & $1.7\pm0.5$  &$0.1\pm0.1$&\multicolumn{1}{c}{--} \\ 
Hadron$+\gamma$ &
$1.4\pm0.7$ &$0.5\pm0.3$ & $0.2\pm0.1$&\multicolumn{1}{c}{--} \\ 
$\pi/K$ Decay$+\gamma$ &
$0.8\pm0.9$ &$0.3\pm0.3$ & $0.1\pm0.1$& \multicolumn{1}{c}{--}\\ 
$b/c$ \  \ Decay$+\gamma$   & 
$0.1\pm0.1$ & $\lt 0.01$  &$\lt 0.01$ & \multicolumn{1}{c}{--}\\ 
\hline
Predicted $\mu_{SM}$ &
$24.9\pm2.4$ & $20.2\pm1.7$ & $7.6\pm0.7$& $5.8\pm0.6$\\  
Observed $N_0$& 
\multicolumn{1}{c}{33} &      
\multicolumn{1}{c}{27} &      
\multicolumn{1}{c}{16} &      
\multicolumn{1}{c}{5} 
\\ 
$P(N\geq N_{0}|\mu_{SM})$ & 
\multicolumn{1}{c}{9.3\%} & 
\multicolumn{1}{c}{10.0\%} & 
\multicolumn{1}{c}{0.7\%} & 
\multicolumn{1}{c}{68.0\%}   
\\ 
\hline\hline
\end{tabular}
\end{center}
\caption{The mean numbers $\mu_{SM}$ of two-body and inclusive multi-body
events predicted by the SM, the number $N_0$ observed, and the
observation likelihood $P(N\geq N_{0}|\mu_{SM})$.  Correlated
uncertainties have been taken into account.}
\label{lg result}
\end{table}

\begin{figure*}[!t]
\begin{minipage}[c]{0.5\textwidth}
\includegraphics[width=\textwidth]{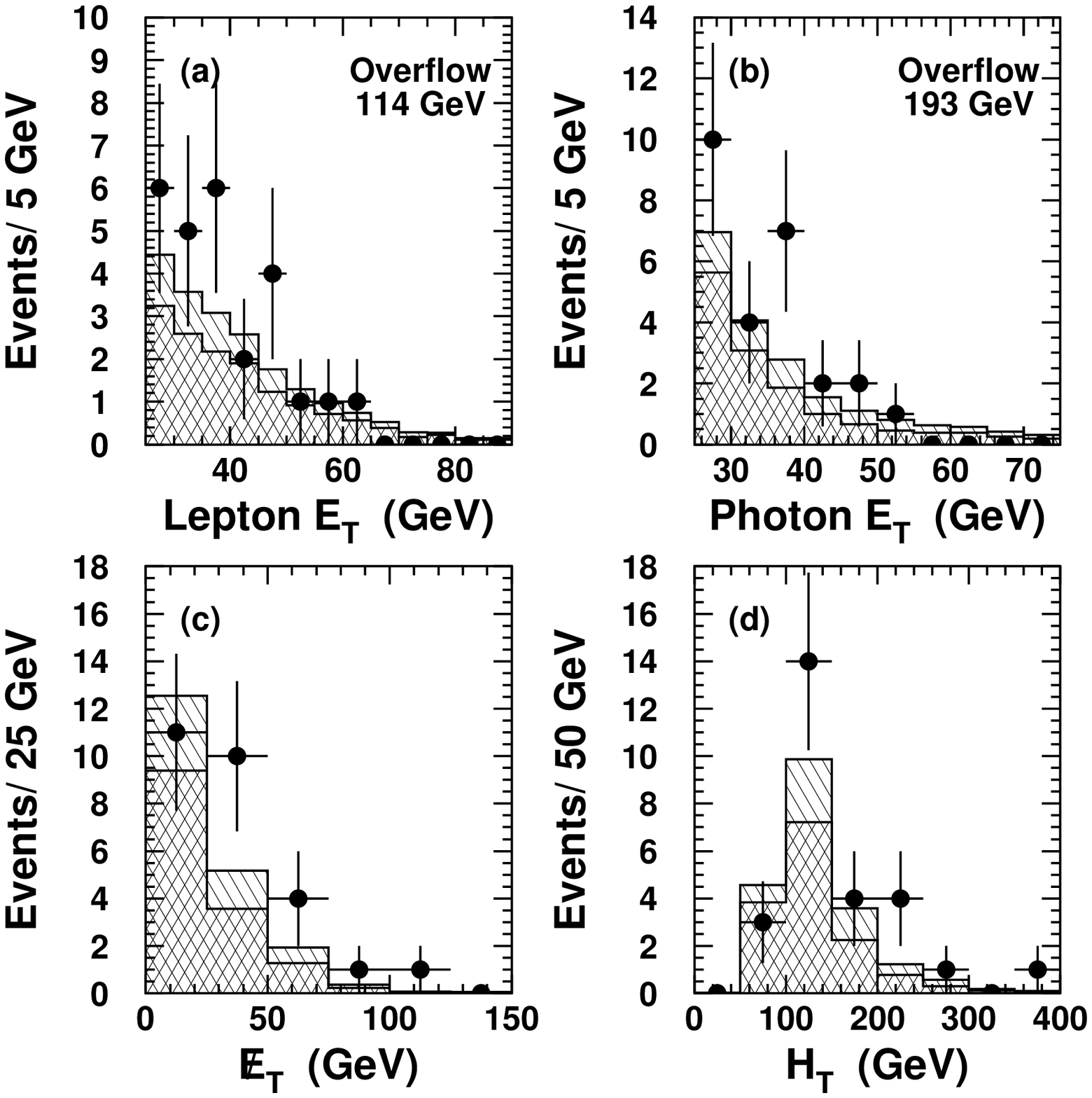}
\end{minipage}%
\begin{minipage}[c]{0.5\textwidth}
\includegraphics[width=\textwidth]{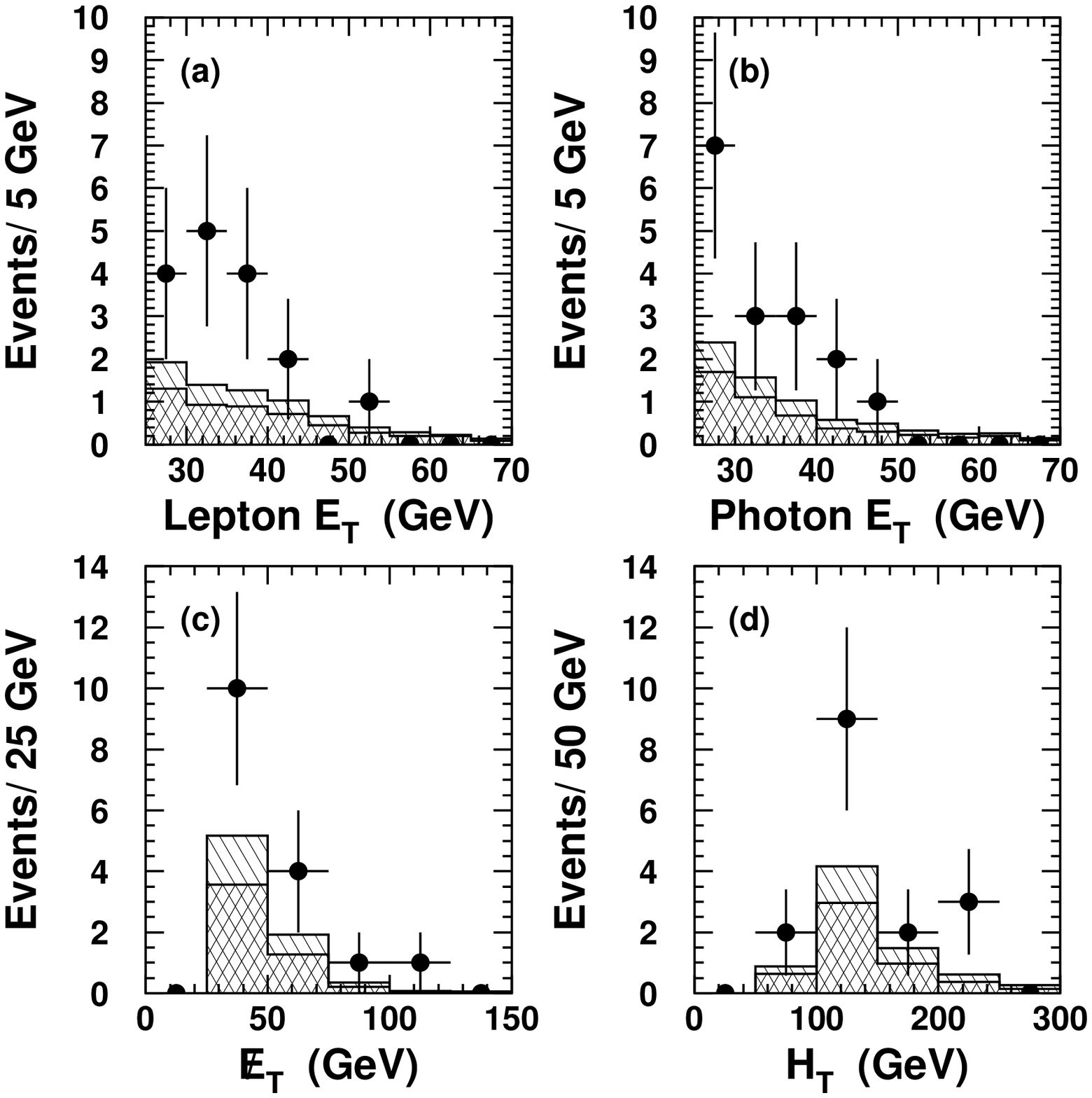}
\end{minipage}
\caption{  
Left: The lepton $\Et$, photon $\Et$, $\mett$, and $\Ht$ of inclusive
multi-body photon-lepton candidates (points) compared to SM 
predictions (single-hatched histograms).  
The cross-hatched histograms show the contribution from SM $W\gamma$
and $Z\gamma$ production. There is one event in the overflow bin 
at 114 GeV in $a$) and one at 193 GeV in $b$). 
Right: The same distributions for the
subset of these events that have $\mett \gt 25\GeV$.}
\label{mblg result}
\end{figure*}

The predicted and observed totals for inclusive multi-body
photon-lepton events are also compared in Table~\ref{lg result}.
About half of the predicted total originates from $Z^{0}\gamma$
production, a quarter from $W\gamma$ production, and the remaining
quarter from particles misidentified as photons or leptons.  The
likelihood of the observed inclusive multi-body total
is 10\%.  The predicted and observed kinematic distributions for these
events are compared in Figure~\ref{mblg result}.  The difference
between the observed and predicted totals can be entirely attributed
to events with $\mett \gt 25\GeV$.  Figure~\ref{mblg result} also
shows the distribution in $\Ht$, the scalar sum of the $\Et$ of all
objects in the event plus the magnitude of $\mett$, a variable
correlated with the production of massive particles~\cite{jeffPRD}.

The predicted and observed totals for multi-body $\ell\gamma\mett$
events are also compared in Table~\ref{lg result}.  For photon-electron
events, requiring $\mett \gt 25\GeV$ suppresses the contributions from
$Z^{0}\gamma$ production and from electrons misidentified as photons,
which have no intrinsic $\mett$, while preserving the contribution
from $W\gamma$ production.  As a result, 57\% of the predicted
$e\gamma\mett$ total arises from $W\gamma$ production, 31\% from jets
misidentified as photons, only 3\% from $Z^{0}\gamma$ production, and
the remaining 9\% from other particles misidentified as photons.  The
observed $e\gamma\mett$ total agrees with the predicted total, with a
25\% probability that the predicted mean of $3.4$ events yields
5 observed events. One of these 5 is the $ee\gamma\gamma\mett$ 
event~\cite{eegg}.

For photon-muon events, requiring $\mett \gt 25\GeV$ does not
completely eliminate the contribution from $Z^{0}\gamma$, for a
second muon with $|\eta| \gt 1.2$ and $\Pt \gt 25\GeV$ can escape 
detection and induce the necessary amount of $\mett$.
The rate at which this occurs is modeled well by the $Z^{0}\gamma$ event
simulation, however, since it is largely a function of the CDF detector
geometric acceptance.  Of the 4.6 multi-body 
events predicted to originate from $Z^{0}\gamma$ production, 2.2
events are predicted to contain a second visible muon, 1.0 events 
are predicted to have less than 25\GeV of $\mett$,  
and only 1.0 events are predicted to pass the 25\GeV $\mett$ selection.
One event is observed with a second 
muon, in agreement with the $Z^{0}\gamma$ prediction. 
The predicted total for multi-body $\mu\gamma\mett$ events consists
of 47\% $W\gamma$ production, 24\% events with jets misidentified as
photons, 23\% $Z^{0}\gamma$ production, and the remaining 7\% from
particles misidentified as muons.    
 
The $\mu\gamma\mett$ event total is higher than predicted (11
observed vs. 4 expected), with an observation likelihood of 
0.54\%; the observation likelihood of the $\ell\gamma\mett$ total
is only slightly higher at 0.72\%~\cite{e_mu_diff}.
The predicted and observed distributions of the kinematic properties
of multi-body $\ell\gamma\mett$ events are compared in
Figure~\ref{mblg result}.  The observed photon $\Et$,
lepton $\Et$, $\mett$, and $\Ht$ distributions are within the range
expected from the SM~\cite{referee_motivation}.

The predicted and observed totals of multi-lepton
events are compared in Table~\ref{lg result}.  Nearly all 
of the predicted total is expected from $Z^{0}\gamma$ production.
Approximately 6 events are expected; 5 events are observed,
including the $ee\gamma\gamma\mett$ event. 
No $e\mu\gamma$ events were expected, and none were observed.

The predicted number of multi-photon events is dominated by
$Z\gamma$ production, for which only  0.01 events
are expected. The single event
observed is the $ee\gamma\gamma\mett$ event, whose (un)likelihood is
described in Ref.~\cite{eegg}.

In conclusion, we have made an \textit{a priori} search for inclusive
photon+lepton production. We find that subsamples of this data set
agree well with their SM prediction, with the possible exception of
$\gamma\ell\mett$. However, an excess of events with 0.7\%
likelihood (equivalent to 2.7 standard deviations for a Gaussian
distribution) in one subsample among the five studied is an
interesting result, but is not a compelling observation of new
physics.  We look forward to more data in the upcoming run of the
Fermilab Tevatron.

\begin{acknowledgments}
We thank the Fermilab staff and the technical staffs of the
participating institutions for their contributions.  We thank U.~Baur
and S.~Mrenna for their critical calculations of the SM $W\gamma$ and
$Z\gamma$ backgrounds used in this analysis.  This work was supported
by the U.S. Department of Energy and National Science Foundation; the
Italian Istituto Nazionale di Fisica Nucleare; the Ministry of
Education, Culture, Sports, Science, and Technology of Japan; the
Natural Sciences and Engineering Research Council of Canada; the
National Science Council of the Republic of China; the Swiss National
Science Foundation; the A. P. Sloan Foundation; the Bundesministerium
fuer Bildung und Forschung, Germany; the Korea Science and Engineering
Foundation (KoSEF); the Korea Research Foundation; and the Comision
Interministerial de Ciencia y Tecnologia, Spain.
                                                                  
\end{acknowledgments}

\end{document}

%% file: run1_revtex4_auth.tex
%
%
\affiliation{Institute of Physics, Academia Sinica, Taipei, Taiwan 11529, 
Republic of China}
\affiliation{Argonne National Laboratory, Argonne, Illinois 60439}
\affiliation{Istituto Nazionale di Fisica Nucleare, University of Bologna,
I-40127 Bologna, Italy}
\affiliation{Brandeis University, Waltham, Massachusetts 02254}
\affiliation{University of California at Davis, Davis, California  95616}
\affiliation{University of California at Los Angeles, Los 
Angeles, California  90024} 
\affiliation{Instituto de Fisica de Cantabria, CSIC-University of Cantabria, 
39005 Santander, Spain}
\affiliation{Carnegie Mellon University, Pittsburgh, PA  15218} 
\affiliation{Enrico Fermi Institute, University of Chicago, Chicago, 
Illinois 60637}
\affiliation{Joint Institute for Nuclear Research, RU-141980 Dubna, Russia}
\affiliation{Duke University, Durham, North Carolina  27708} 
\affiliation{Fermi National Accelerator Laboratory, Batavia, Illinois 
60510}
\affiliation{University of Florida, Gainesville, Florida 32611}
\affiliation{Laboratori Nazionali di Frascati, Istituto Nazionale di Fisica
               Nucleare, I-00044 Frascati, Italy}
\affiliation{University of Geneva, CH-1211 Geneva 4, Switzerland} 
\affiliation{Glasgow University, Glasgow G12 8QQ, United Kingdom}
\affiliation{Harvard University, Cambridge, Massachusetts 02138} 
\affiliation{Hiroshima University, Higashi-Hiroshima 724, Japan}
\affiliation{University of Illinois, Urbana, Illinois 61801}
\affiliation{The Johns Hopkins University, Baltimore, Maryland 21218}
\affiliation{Institut f\"{u}r Experimentelle Kernphysik, 
Universit\"{a}t Karlsruhe, 76128 Karlsruhe, Germany}
\affiliation{Center for High Energy Physics: Kyungpook National
University, Taegu 702-701; Seoul National University, Seoul 151-742; and
SungKyunKwan University, Suwon 440-746; Korea}
\affiliation{High Energy Accelerator Research Organization (KEK), Tsukuba, 
Ibaraki 305, Japan}
\affiliation{Ernest Orlando Lawrence Berkeley National Laboratory, 
Berkeley, California 94720}
\affiliation{Massachusetts Institute of Technology, Cambridge,
Massachusetts  02139} 
\affiliation{University of Michigan, Ann Arbor, Michigan 48109}
\affiliation{Michigan State University, East Lansing, Michigan  48824}
\affiliation{University of New Mexico, Albuquerque, New Mexico 87131} 
\affiliation{Northwestern University, Evanston, Illinois  60208} 
\affiliation{The Ohio State University, Columbus, Ohio  43210}
\affiliation{Osaka City University, Osaka 588, Japan} 
\affiliation{University of Oxford, Oxford OX1 3RH, United Kingdom} 
\affiliation{Universita di Padova, Istituto Nazionale di Fisica 
          Nucleare, Sezione di Padova, I-35131 Padova, Italy}
\affiliation{University of Pennsylvania, Philadelphia, 
        Pennsylvania 19104}
\affiliation{Istituto Nazionale di Fisica Nucleare, University and Scuola
               Normale Superiore of Pisa, I-56100 Pisa, Italy} 
\affiliation{University of Pittsburgh, Pittsburgh, Pennsylvania 15260} 
\affiliation{Purdue University, West Lafayette, Indiana 47907}
\affiliation{University of Rochester, Rochester, New York 14627}
\affiliation{Rockefeller University, New York, New York 10021}
\affiliation{Rutgers University, Piscataway, New Jersey 08855}
\affiliation{Texas A\&M University, College Station, Texas 77843}
\affiliation{Texas Tech University, Lubbock, Texas 79409}
\affiliation{Institute of Particle Physics, University of Toronto, Toronto
M5S 1A7, Canada}
\affiliation{Istituto Nazionale di Fisica Nucleare, University of 
Trieste/Udine, Italy}
\affiliation{University of Tsukuba, Tsukuba, Ibaraki 305, Japan}
\affiliation{Tufts University, Medford, Massachusetts 02155}
\affiliation{Waseda University, Tokyo 169, Japan}
\affiliation{University of Wisconsin, Madison, Wisconsin 53706}
\affiliation{Yale University, New Haven, Connecticut 06520}

\author{D.~Acosta}
\affiliation{University of Florida, Gainesville, Florida 32611}

\author{T.~Affolder}
\affiliation{Ernest Orlando Lawrence Berkeley National Laboratory, 
Berkeley, California 94720}

\author{H.~Akimoto}
\affiliation{Waseda University, Tokyo 169, Japan}

\author{M.~G.~Albrow}
\affiliation{Fermi National Accelerator Laboratory, Batavia, Illinois 
60510}

\author{D.~Ambrose}
\affiliation{University of Pennsylvania, Philadelphia, 
        Pennsylvania 19104}

\author{D.~Amidei}
\affiliation{University of Michigan, Ann Arbor, Michigan 48109}

\author{K.~Anikeev}
\affiliation{Massachusetts Institute of Technology, Cambridge,
Massachusetts  02139} 

\author{J.~Antos}
\affiliation{Institute of Physics, Academia Sinica, Taipei, Taiwan 11529, 
Republic of China}

\author{G.~Apollinari}
\affiliation{Fermi National Accelerator Laboratory, Batavia, Illinois 
60510}

\author{T.~Arisawa}
\affiliation{Waseda University, Tokyo 169, Japan}

% ...


%% file: prl_shortened.bbl
\begin{thebibliography}{99}
\bibitem{CDFcoo}  The CDF coordinate system of 
$r$, $\varphi$, and $z$ is cylindrical, with the $z$-axis along the
proton beam.  The pseudorapidity is $\eta = -\ln(\tan(\theta/2))$.
\bibitem{EtPt} The transverse momentum is defined as $\Pt =
p\sin\theta$; the transverse energy is defined as $\Et = E\sin\theta$.
Missing transverse energy is defined as the vector opposite to the
vector sum of the transverse energies of all objects in an event,
$\vec{{\rm \not\! E}_{\rm T}} = -\Sigma\vec{\Et}$. We use the 
convention that `momentum' refers to $pc$ and `mass' to $mc^2$, so that energy,
momentum, and mass are all measured in GeV.
\bibitem{delta_mu}
The uncertainty in $\mu_{SM}$ is the standard deviation of a
large ensemble of calculations, and takes into account correlated
uncertainties.
\bibitem{e_mu_diff} The observed
muon-electron difference is statistically insignificant; the probability
that 16 events divide up into two categories 
as asymmetrically as 11 to 5 or higher is greater than 8\%.
\end{thebibliography}
